# Key Technologies for the Wide Field Infrared Survey Telescope Coronagraph Instrument

A white paper submitted in response to the National Academies of Science, Engineering and Medicine's Call on **Exoplanet Science Strategy** (March 2018)[1].

Vanessa P. Bailey
Jet Propulsion Laboratory, California Institute of Technology
vanessa.bailey@jpl.nasa.gov, +1 818-354-2034

Lee Armus *(1/2)*, Bala Balasubramanian *(3/1)*, Pierre Baudoz *(4)*, Andrea Bellini *(5)*, Dominic Benford *(6)*, Bruce Berriman *(1/7)*, Aparna Bhattacharya *(8)*, Anthony Boccaletti *(4/9)*, Eric Cady *(3/1)*, Sebastiano Calchi Novati *(1/2)*, Kenneth Carpenter *(8)*, David Ciardi *(1/7)*, Brendan Crill *(3/1)*, William Danchi *(8)*, John Debes *(5)*, Richard Demers *(3/1)*, Kjetil Dohlen *(10/9)*, Robert Effinger *(3/1)*, Marc Ferrari *(10)*, Margaret Frerking *(3/1)*, Dawn Gelino *(1/7)*, Julien Girard *(5)*, Kevin Grady *(8)*, Tyler Groff *(8)*, Leon Harding *(3/1)*, George Helou *(1)*, Avenhaus Henning *(11)*, Markus Janson *(12)*, Jason Kalirai *(5)*, Stephen Kane *(13)*, N. Jeremy Kasdin *(14)*, Matthew Kenworthy *(15)*, Brian Kern *(3/1)*, John Krist *(3/1)*, Jeffrey Kruk *(8)*, Anne Marie Lagrange *(16/9)*, Seppo Laine *(1/2)*, Maud Langlois *(9/17)*, Hervé Le Coroller *(10/9)*, Chris Lindensmith *(3/1)*, Patrick Lowrance *(1/18)*, Anne-Lise Maire *(11)*, Sangeeta Malhotra *(8)*, Avi Mandell *(8)*, Michael McElwain *(8)*, Camilo Mejia Prada *(3/1)*, Bertrand Mennesson *(3/1)*, Tiffany Meshkat *(1/2)*, Dwight Moody *(3/1)*, Patrick Morrissey *(3/1)*, Leonidas Moustakas *(3/1)*, Mamadou N'Diaye *(10)*, Bijan Nemati *(19)*, Charley Noecker *(3/1)*, Roberta Paladini *(1/2)*, Marshall Perrin *(5)*, Ilya Poberezhskiy *(3/1)*, Marc Postman *(5)*, Laurent Pueyo *(5)*, Solange Ramirez *(1/2)*, Clément Ranc *(8)*, Jason Rhodes *(3/1)*, A.J.E. Riggs *(3/1)*, Maxime Rizzo *(8)*, Aki Roberge *(8)*, Daniel Rouan *(20/21)*, Joshua Schlieder *(8)*, Byoung-Joon Seo *(3/1)*, Stuart Shaklan *(3/1)*, Fang Shi *(3/1)*, Rémi Soummer *(5)*, David Spergel *(14)*, Karl Stapelfeldt *(3/1)*, Christopher Stark *(5)*, Motohide Tamura *(22)*, Hong Tang *(3/1)*, John Trauger *(3/1)*, Margaret Turnbull *(23)*, Roeland van der Marel *(5)*, Arthur Vigan *(10/9)*, Benjamin Williams *(24)*, Edward J. Wollack *(8)*, Marie Ygouf *(1/2)*, Feng Zhao *(3/1)*, Hanying Zhou *(3/1)*, and Neil Zimmerman *(8)*

*1. Caltech, 2. IPAC, 3. JPL, 4. Paris Observatory, 5. STScI, 6. NASA, 7. IPAC-NExScI, 8. GSFC, 9. CNRS, 10. LAM, 11. MPIA, 12. Stockholm University, 13. UC Riverside, 14. Princeton, 15. Leiden Observatory, 16. IPAG, 17. CRAL, 18. IPAC-Spitzer, 19. Univ. of Alabama – Huntsville, 20. LESIA, 21. Obs. de Paris, 22. NAOJ, 23. Global Science Institute, 24. University of Washington*

---

[1] This whitepaper was submitted to the Exoplanet Science Strategy call in March 2018 and is presented here without modification. An updated version of Fig. 2 and associated detailed description can be found in ref (21).

# 1 Introduction

The Wide Field Infrared Survey Telescope (WFIRST) Coronagraph Instrument (CGI) is a high-contrast imager and integral field spectrograph that will enable the study of exoplanets and circumstellar disks at visible wavelengths [Mennesson and Kasdin white papers]. Ground-based high-contrast instrumentation is fundamentally limited to flux ratios of $10^{7-8}$ at small working angles, even under optimistic assumptions for 30m-class telescopes (1; 2). There is a strong scientific driver for better performance, particularly at visible wavelengths [Seager white paper]. Future flagship mission concepts aim to image Earth analogues with visible light flux ratios $>10^{10}$ [Crill, HabEx, and LUVIOR white papers; (3)]. CGI is a critical intermediate step toward that goal, with a predicted $10^{8-9}$ flux ratio capability. CGI achieves that capability through improvements over current ground and space systems in several areas:

- Hardware: space-qualified (TRL9) deformable mirrors, detectors, and coronagraphs
- Algorithms: wavefront sensing and control; post-processing of integral field spectrograph, polarimetric, and extended object data
- Validation of telescope and instrument models at high accuracy and precision

This white paper describes the current status of key technologies and presents ways in which performance is likely to evolve as the CGI design matures. WFIRST is now in Phase A; this paper is not intended as a definitive document on the final instrument configuration or performance.

# 2 Key CGI components

- Two science cameras (cannot be used simultaneously):
    - Imager: 10" FOV; direct imaging or polarimetry; 10% bandpass filters
    - Integral Field Spectrograph (IFS): 2" FOV; R~50 spectrum; 18% bandpass filters.
    - Electron Multiplying CCDs (EMCCDs) for improved signal-to-noise on faint objects.
- Visible to very near infrared wavelengths:
    - 10% bandwidth: 575nm and 825nm; 18% bandwidth: 660nm and 760nm
    - 1% bandwidth Hα filter for IFS calibration and imaging is under consideration.
- Starlight suppression with interchangeable coronagraphic masks[2]:
    - Hybrid Lyot Coronagraph (HLC): 360dgr FOV, 3-9λ/D, optimized for imaging.
    - Shaped Pupil Coronagraph (SPC) "bowtie": 2 x 65dgr FOV, 3-9λ/D, optimized for the broader IFS bandpasses.
    - SPC "disk": 360dgr FOV, 6.5-20λ/D, optimized for imaging.
- Wavefront sensing and control at unprecedented levels of precision:
    - Dedicated Low Order Wavefront Sensor (LOWFS) for Zernike modes 2-11.
    - High Order Wavefront Sensing (HOWFS) using science camera images.
    - Two high-actuator count deformable mirrors (DMs) for phase and amplitude control.

The current budget allows for fully commissioning three observing modes: 575nm/HLC/imaging, 760nm/SPC bowtie[3]/IFS, and 825nm/SPC disk/imaging. These modes will be tested with CGI flight hardware and software. Other combinations of filters and coronagraphic masks are possible and will be exercised at the JPL WFIRST CGI engineering testbed, though they will not be fully tested with flight hardware and software prior to launch, due to CGI Integration and Test schedule and budget constraints.

---

[2] The use of an external occulting "starshade" with WFIRST is under consideration [Ziemer WP], pending guidance from the next Decadal Survey.
[3] Only one SPC bowtie orientation is included in the current baseline. Installation of the remaining two SPC bowtie masks, without full pre-flight commissioning, is under consideration.

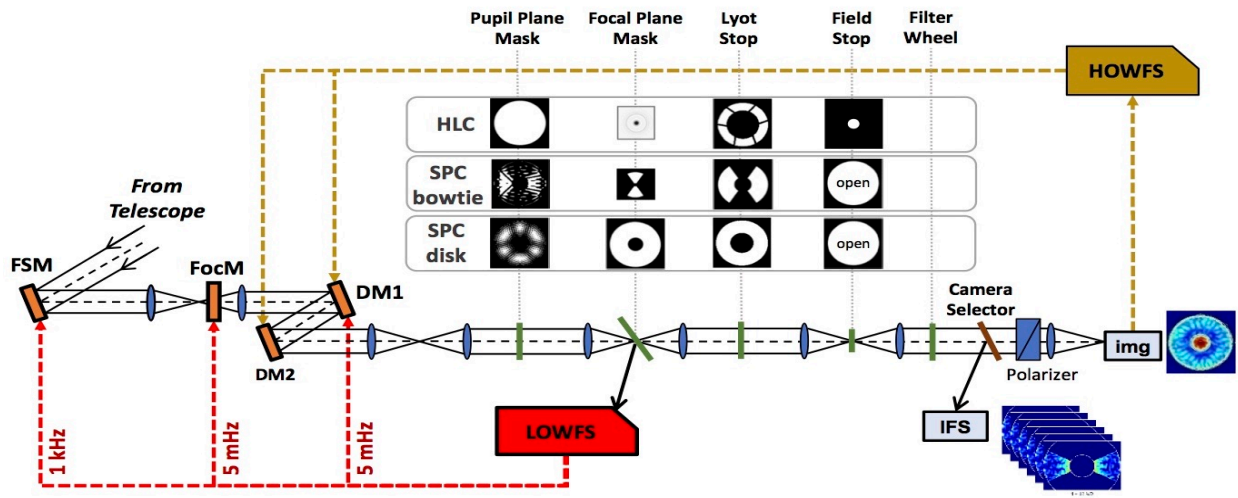

Figure 1: CGI schematic diagram.

## 3 Coronagraph Designs

CGI has chosen two families of coronagraphs, Hybrid Lyot and Reflective Shaped Pupil, on the basis of their maturity, expected performance with the WFIRST obscured pupil, and low sensitivity to aberrations (4; 5). New fabrication techniques have been implemented to address the tight optical tolerances (6; 7). The designs must be robust against effects that were not significant at the contrast levels achieved by previous-generation coronagraphs. For example, accommodating the secondary mirror support struts pushes designs to more difficult trades between performance metrics such as IWA, throughput, bandwidth, contrast, field-of-view, and aberration sensitivities, relative to designs for unobscured apertures. Additionally, polarization-dependent aberrations and telescope tip/tilt jitter limit starlight suppression at small working angles; ongoing work is evaluating soft-edge focal plane masks to reduce sensitivity to these effects. Future flagship mission concepts are already learning from CGI experience in areas including: coronagraph designs for complex apertures, mirror coatings to minimize polarization-dependent aberrations, and lower-vibration spacecraft pointing control systems.

## 4 Wavefront Sensing and Control

State of the art ground-based adaptive optics systems control the incoming wavefront to tens of nanometers RMS. CGI must stabilize the wavefront to tens of picometers RMS, and future exo-Earth imaging missions aim for <10pm RMS. This level of wavefront control is infeasible even on future 30m-class ground-based facilities [eg: (1; 2)]. The surest path to imaging Jovian and Earth analogues in reflected visible light is to move to a stable, space-based platform.

### 4.1 Low Order Aberrations

Tip/tilt errors will originate from slow (sub-Hz) observatory pointing drift and from structural vibrations excited by the telescope reaction wheels (1-100Hz). Longer timescale thermal drifts in the spacecraft will be the primary contributors to errors in other low order modes. To compensate, CGI will have a dedicated Low Order Wavefront Sensor and Control system (LOWFS/C; (8)) for Zernike modes 2-11. The Zernike phase contrast wavefront sensor uses rejected starlight reflected by the coronagraph occulting masks. A fast steering mirror will correct tip/tilt jitter at frequencies ≲20Hz; other modes will be corrected at 5mHz with a combination of a dedicated focus corrector and the DMs. LOWFS images will be downlinked at full frame rate (1kHz) for use in post-processing and in telescope model validation to inform future missions.

Lab tests of the LOWFS/C have shown promising performance, with several important tests remaining. On bright sources (V = -5), with disturbances approximating the anticipated telescope error power spectrum, the engineering testbed has demonstrated tip/tilt control to better than 0.5mas RMS and focus sensing at 10pm accuracy with closed-loop rejection of 20dB/decade (9). Future work will verify sensing and control of other modes. In the coming year, fainter (V=5) sources will be tested; models predict sensing on V<6 stars will be photon noise-limited. Finally, we note that LOWFS optical alignment tolerances require thermal regulation to better than 0.1K over tens of hours; future observatories will need even more stringent controls.

## 4.2 Higher Order Aberrations

For higher-order modes, CGI will employ focal-plane wavefront sensing, using science camera images themselves; corrections will be applied to the DMs. The baseline wavefront sensing and correction scheme is pairwise probing and electric field conjugation (10; 11). This scheme has been demonstrated in the engineering testbed for both the SPC bowtie and HLC modes (12; 13), achieving contrasts below $5\times10^{-9}$ and $1\times10^{-9}$, respectively. More recent tests have demonstrated that the control scheme may be used while flight-like tip/tilt jitter and low-order drift are injected upstream and corrected by the LOWFS/C (9). Future tests will verify performance of newer HLC and SPC designs on fainter (V=5) sources and with polarization-dependent low-order aberrations. One remaining challenge is instrument-model mismatch, which slows wavefront correction convergence and limits achievable contrast. Work is ongoing to improve characterization of as-built coronagraph masks, optics, and DM actuator influence functions. New algorithms that update the instrument model in situ, using feedback from previous iterations, are another promising path for improvement (14).

## 4.3 Deformable Mirrors

Achieving an annular dark hole requires two deformable mirrors to correct both the amplitude and phase components of the complex electric field (15; 16). WFIRST will place one DM at a pupil image plane and a second out-of-plane. The DMs, from Xinetics, each have 48x48 electrostrictive actuators and were chosen for their relatively high level of technology readiness.

Because CGI high-order wavefront sensing is photon-starved, the DMs will be tuned on a bright reference star before slewing to the science target. The DMs are required to remain stable throughout a ~10hr science sequence, without active control of dynamic optical aberrations higher than the Zernike modes sensed by the LOWFS. The CGI DMs must therefore be calibrated and stabilized to levels far exceeding those of ground-based applications. A three-pronged approach of precise thermal control (~3mK stability), improved commanding (accounting for long-timescale actuator settling), and actuator-by-actuator calibration (gains, stability, and surface influence profiles) is the subject of ongoing effort.

## 5 Detector development

Both science cameras and the LOWFS will use Electron Multiplying CCDs (EMCCDs) (17). These devices amplify the photoelectron signal at readout, giving superior signal-to-noise in the very low flux regime, relative to classical CCDs. EMCCDs have not flown in space-based science instruments; information from CGI on detector performance, systematics, and degradation over multi-year timescales is essential for formulating future missions. The imager and IFS detectors will typically operate in high-gain "photon counting" mode, where the frame time is set so that each pixel has at most one photoelectron. This bypasses the effect of EM gain uncertainty (the "extra noise factor") and maintains the effective quantum efficiency of the detector. Frame times <100s also minimize contamination from cosmic rays, which are a significant noise source.

Ongoing work is addressing several hardware and software challenges. Detector degradation from high energy particle damage is a concern for operations beyond the initial 18 month technology demonstration period. Damaged pixels can have both higher dark current and more "charge traps;" the latter lead to lower effective detector quantum efficiency. Additionally, large signals, such as cosmic rays, induce "tails" of charge in adjacent pixels during high-gain read; it is estimated that this effect could contaminate 10-20% of pixels in photon counting mode. Currently, these effects are the limiting factors for IFS sensitivity. JPL is iterating with Teledyne-e2v to produce new devices that address each of these effects. New devices will be tested prior to PDR, including a radiation test campaign. Additionally, CGI will build on post-processing strategies, pioneered on HST, for mitigating the impact of cosmic rays and charge traps (18).

## 6 Integral Field Spectroscopy

The IFS will enable the first demonstrations of atmosphere spectral retrieval at very high (>$10^8$) flux ratios, providing system-level scientific operations experience to benefit future flagship missions. The R = 50 spectra will also support the mission objective of validating integrated observatory and instrument models by capturing the chromatic behavior of the speckle noise floor in a new contrast regime. The prioritized spectroscopy demonstration filter is an 18% bandpass centered at 760nm; however, the IFS can operate in a $\Delta\lambda/\lambda$ = 20% instantaneous bandpass anywhere in the wavelength range 600—1000 nm.

## 7 Post-processing

Lessons learned from existing high-contrast instrumentation provide a solid foundation for processing CGI data. Techniques such as angular and reference differential imaging (ADI, RDI) will still be critical. Current baseline observing scenarios include both two-roll imaging (limited by telescope sun angle constraints) as well as target-reference chopping. In this new very high-contrast regime, speckles are affected by both phase and amplitude aberrations. Hence, existing tools that hinge on phase-only dependence must be updated (19). Additionally, aberrations will be polarization-dependent; therefore, polarimetry will only be possible for targets that are much brighter than the speckle floor, unless new algorithms are invented.

Algorithm development with realistic simulated images will be a high priority during Phase B. Challenges include: spectral retrieval in the presence of detector artifacts and other systematics, as well as reconstruction of extended, low surface brightness features in both polarized and unpolarized light. Incorporating LOWFS and telescope telemetry, including tracking the star position under the coronagraph and creating physically motivated modal basis sets for PSF subtraction, is an area of active research.

## 8 Performance Prediction

Validating performance of CGI pre- and post-launch will heavily rely on comparison to simulation. Good agreement has been demonstrated for lab tests of the SPC/IFS mode in several metrics: raw contrast, HOWFS/C convergence rate, and key contrast stability sensitivities (20). Future work will improve model predictions for other operational modes, for new wavefront control schemes, and for other, more flight-like, configurations.

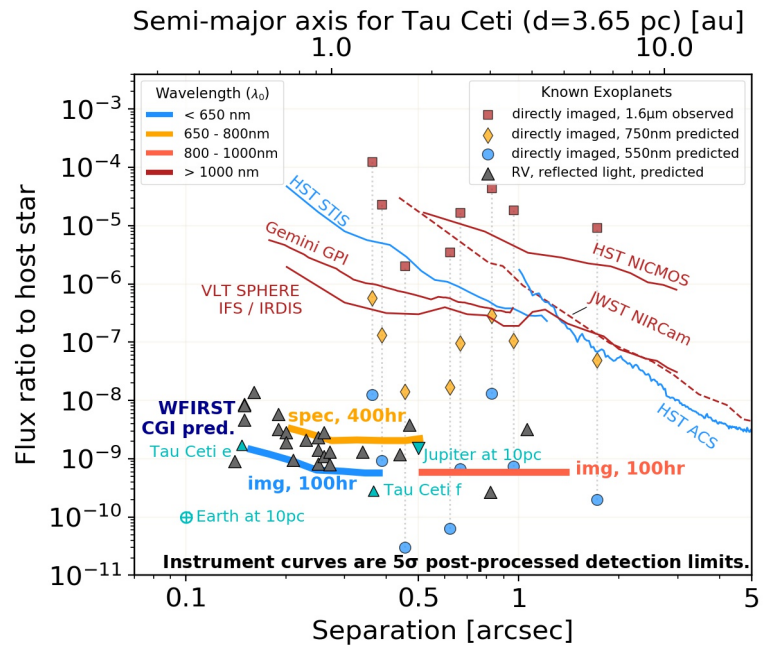

Figure 2: Predicted CGI performance on a V=5 star, in the context of known giant planets and current instrumentation. See (21) for a description of assumptions and data.

Figure 2 presents model predictions for CGI imaging and IFS sensitivity, based on current lab-validated performance of the coronagraphs, wavefront control, and detectors[4]. Telescope vibration, aberration, and thermal environments predicted by integrated modeling are assumed; model uncertainty factors of unity are used throughout. Shot noise from the planet, residual stellar PSF, zodi, and exo-zodi is included. Until post-processing algorithms are more mature, a conservative scenario is used: simple roll/reference image subtraction with an additional factor of two gain from the application of all other post-processing algorithms.

---

[4] An exception was the SPC "disk" mask, for which end-to-end performance modeling was less mature at the time of NAS submission. An updated performance prediction is available at (21).